# Magnetic Doping and Kondo Effect in $Bi_2Se_3$ Nanoribbons

*Judy J. Cha[1], James R. Williams[2], Desheng Kong[1], Stefan Meister[1], Hailin Peng[1, §], Andrew J. Bestwick[2], Patrick Gallagher[2], David Goldhaber-Gordon[2], and Yi Cui[1,*]*

[1]Department of Materials Science and Engineering, Stanford University, Stanford, CA 94305, USA

[2]Department of Physics, Stanford University, Stanford, CA 94305, USA

[§]Present address: College of Chemistry and Molecular Engineering, Peking University, Beijing 100871, P. R. China.

*To whom correspondence should be addressed. E-mail: yicui@stanford.edu.

*A simple surface band structure and a large bulk band gap have allowed $Bi_2Se_3$ to become a reference material for the newly discovered three-dimensional topological insulators, which exhibit topologically-protected conducting surface states that reside inside the bulk band gap. Studying topological insulators such as $Bi_2Se_3$ in nanostructures is advantageous because of the high surface-to-volume ratio, which enhances effects from the surface states; recently reported Aharonov-Bohm oscillation in topological insulator nanoribbons by some of us is a good example. Theoretically, introducing magnetic impurities in topological insulators is predicted to open a small gap in the surface states by breaking time-reversal symmetry. Here, we present synthesis of magnetically-doped $Bi_2Se_3$ nanoribbons by vapor-liquid-solid growth using magnetic metal thin films as catalysts.*

*Although the doping concentration is less than ~ 2%, low-temperature transport measurements of the Fe-doped $Bi_2Se_3$ nanoribbon devices show a clear Kondo effect at temperatures below 30 K, confirming the presence of magnetic impurities in the $Bi_2Se_3$ nanoribbons. The capability to dope topological insulator nanostructures magnetically opens up exciting opportunities for spintronics.*

Following the discovery of the quantum spin Hall insulator state in two dimensions, where a robust edge conducting state coexists with insulating bulk,[1-4] a three-dimensional (3D) topological insulator phase has attracted much interest. 3D topological insulators are a new state of quantum matter,[5-7] whose key property is the topologically protected, conducting surface states that reside inside a bulk band gap. Such states possess properties that are ideal for dissipation-less electronics and spintronics applications. The 3D topological surface states were first observed in $Bi_{1-x}Sb_x$ alloy with a complicated surface band structure.[8] Among the predicted and experimentally verified topological insulators,[9-13] $Bi_2Se_3$ serves as a reference material because of its simple band structure and the large bulk band gap of ~ 0.3 eV.[9] The signature of the surface states, a Dirac-like band structure, and the robustness of the surface states have been confirmed in bulk samples with the angle-resolved photo-emission spectroscopy (ARPES) and scanning tunneling spectroscopy methods.[11, 13] Compared to bulk $Bi_2Se_3$ samples, nano-structured $Bi_2Se_3$ offers an attractive alternative for studying the surface states due to the high surface-to-volume ratio, which greatly enhances the visibility of the surface states in transport measurements. Thin films of $Bi_2Se_3$ grown by molecular beam epitaxy show the surface states very clearly, observed by ARPES.[14, 15] Recently, Aharonov-Bohm oscillations in $Bi_2Se_3$ nanoribbons were observed by some of the present authors to prove unambiguously the existence of surface conducting states in transport measurements.[16]

The surface states of topological insulators are robustly protected by time-reversal symmetry. Breaking time-reversal symmetry by doping topological insulators with magnetic impurities is predicted to open a gap in the otherwise gapless surface states.[17] Here, we propose a general approach to *in-situ* dope $Bi_2Se_3$ nanoribbons with magnetic impurities. By choosing magnetic metal alloys as a catalyst during the vapor-liquid-solid (VLS) growth,[18] we show that the $Bi_2Se_3$ nanoribbons can be magnetically doped and exhibit Kondo effect at temperatures below ~ 30 K.

Magnetically doped nanoribbons were synthesized in a horizontal tube furnace via VLS mechanism. Different from the Au-catalyzed growth of $Bi_2Se_3$ nanoribbons reported recently,[16, 19] Ni-Au and Fe-Au thin films were used both as catalysts and as a source of magnetic dopants. Atom probe tomographic reconstructions and careful transport measurements on VLS-grown Si nanowires using Au catalysts reveal that some Au atoms are unavoidably incorporated into the Si nanowires during growth.[20] These observations suggest that, during VLS growth, nanowires and nanoribbons can be doped by the metal catalysts.[21, 22] Ni-Au and Fe-Au thin films were prepared by sputtering Ni (2 nm)/Au (5 nm) or e-beam evaporating Fe (2 nm)/Au (5 nm) on clean Si <100> substrates at the base pressure of ~ $7 \times 10^{-7}$ Torr. The Au layer serves as a capping layer to prevent oxidation of Ni and Fe, and it also promotes a long nanoribbon growth. $Bi_2Se_3$ nanoribbons were grown in the tube furnace by heating a source material ($Bi_2Se_3$ powder from Alfa Aesar) at the center to 530 °C in the presence of a Ar carrier gas, flowing at 135 ~ 145 standard cubic centimeters per minute, and maintaining the temperature and the Ar flow rate for 1 to 3 hours. The Si substrates with the catalyst thin films were placed at the downstream side of the tube furnace where the local temperature ranged between 350 °C and 450 °C. The temperature profile of the tube furnace has been reported previously.[23] In this temperature range, ~ 2% of Ni and ~ 8% of Fe are expected in Au based on thermodynamic-equilibrium phase diagrams.[24, 25] Scanning electron microscopy (SEM) images of the $Bi_2Se_3$ nanoribbons grown using the Fe-Au catalytic thin film are shown in Fig. 1. Two distinct growth products can be found: wide nanoribbons (Fig. 1(a)) and narrow, long nanoribbons (Fig. 1(b)). Both types of nanoribbons show smooth surfaces. As evident in Fig. 1(b), metal particles are observed at the ends of the nanoribbons, which suggests the VLS growth mechanism for these nanoribbons.

Transmission electron microscopy (TEM) studies confirm that the Ni-Au- and Fe-Au-catalyzed growth products are high-quality, single-crystalline $Bi_2Se_3$ nanoribbons. Figure 2 shows TEM images and chemical maps of the as-grown $Bi_2Se_3$ nanoribbons. For TEM characterizations, 200 kV Tecnai F20 equipped with an energy dispersive X-ray (EDX) spectrometer was used. Figure 2(a) and 2(l) show $Bi_2Se_3$ nanoribbons grown using Ni-Au and Fe-Au thin films respectively, where catalyst particles are observed at the end of the nanoribbons. Selected-area electron diffraction (insets in Fig. 2(a) and 2(l)) shows the expected spot patterns with hexagonal symmetry, verifying that $Bi_2Se_3$ nanoribbons are single-crystalline with the correct rhombohedral crystal structure in space group $D^5_{3d}$ (R-3m). High-resolution TEM images (Fig. 2(f) and 2(m)) of the ribbons shown in Fig. 2(a) and 2(l) confirm the expected lattice spacing of ~ 0.21 nm. We also obtained EDX spectra from the ribbons, which show the atomic ratio of Bi to Se to be 2 to 3 within the accuracy of the measurements.

For our proposed *in-situ* magnetic doping via the magnetic metal catalysts to work, it is necessary that magnetic atoms be incorporated into the ribbons during growth. To investigate this, EDX spectra and scanning EDX maps were acquired from the $Bi_2Se_3$ nanoribbons and the catalyst particles at the end of the ribbons. Unsurprisingly, neither Fe nor Ni signals were detected when EDX spectra were obtained from regions that contain only the ribbons, indicating that the incorporation of these magnetic impurities into the ribbons is small, below the detection limit of EDX ( < ~ 2 %). Hence, we focus on the composition of the catalyst particles observed at the end of the ribbons, since if Ni or Fe is present in these particles these magnetic atoms are almost certain to be incorporated into the growing ribbon at some concentration. Usually, when EDX spectra are acquired from the catalyst particles, Au peaks are the dominant feature in the spectra while Fe or Ni peaks are very small as expected from the equilibrium phase diagrams of Ni-Au and Fe-Au. However, we sometimes observe a distinct phase-separation phenomenon in the composite catalyst particles: Au and Fe-Se or Au and Ni-Se. Figure 1(b-e) show Au, Fe, Se, and Bi elemental maps of the ribbon shown in Fig. 1(a); the elemental maps were obtained from a 32x60 pixel EDX scan acquired from the boxed region in scanning TEM mode and each EDX spectrum was acquired for 500 ms. We see that the round metal particle on the right side of the ribbon is Au while the metal particle at the end of the ribbon is Fe-Se. A similar phase-separation is also observed in ribbons grown with Ni-Au thin films. Figure 1(h-k) show Au, Ni, Se, and Bi elemental maps obtained from a 50x70 pixel EDX scan that is acquired from the ribbon shown in 1(g). From the chemical maps, a clear boundary is observed in the metal particle where Au and Ni-Se are phase-separated. In the Ni-Au case, we note that we also observe nanoribbons with only Ni-Se metal particles without Au. The phase separation between Au and Fe or Ni is understandable due to the low solubility of Fe or Ni in Au. Nevertheless, it is interesting to note that both Fe and Ni react with Se to form compound catalyst particles. In summary, we show that the catalyst particles at the end of the nanoribbons contain Fe or Ni based on EDX data, strongly suggesting that some of these magnetic atoms are incorporated in the ribbons as dopants.

Although the VLS growth process with magnetic-alloy catalyst particles suggests that the ribbons are magnetically doped, definitive evidence is required to prove the presence of magnetic dopants in the nanoribbons. To confirm this and to examine how the magnetic dopants affect the electrical properties of the $Bi_2Se_3$ nanoribbons, careful transport measurements in the low temperature regime were carried out, focusing on Fe-doped nanoribbons. To define electrical contacts, a standard electron-beam lithography technique was employed and Cr/Au (typically 7 nm/190 nm) was e-beam evaporated at a base pressure of ~ 8 x $10^{-7}$ Torr.

Temperature-dependent resistance measurements of Fe-doped $Bi_2Se_3$ nanoribbon devices, as well as several devices without magnetic doping, are shown in Fig. 3. For the resistance measurements (Quantum Design PPMS-7 instrument and an Oxford Heliox He3 cryogenic refrigerator), a four-point configuration was used where the voltage drop between the two inner electrodes was measured while flowing a fixed current through the two outer electrodes (Fig. 3(a) inset). The four-point resistance is denoted as $R_{4pt}$. For $Bi_2Se_3$ nanoribbons with no magnetic doping, the resistance monotonically decreases with decreasing temperature, saturating at T ~ 30 K (Fig. 3(a)) and remaining flat down to 2 K, the minimum temperature reached during these measurements. This typical metallic behavior is expected since the carrier concentration in nanoribbons is high, which is caused by materials defects such as vacancies (no intentional doping is added, so below we refer to these samples as "un-doped"). Fig. 3(a) inset is an SEM image of the un-doped nanoribbon device that corresponds to the curve in solid black. In contrast to the un-doped case, the resistance of the Fe-doped $Bi_2Se_3$ nanoribbons increases with decreasing temperature below a certain temperature. Fig. 3(b) plots typical $R_{4pt}$ vs. T curves measured from two Fe-doped $Bi_2Se_3$ nanoribbon devices, showing the clear increase in $R_{4pt}$ below ~ 30K. The rise in resistance in each device is 60 Ω (dashed blue) and 40 Ω (solid black) respectively. To find out what functional form $R_{4pt}$ takes as a function of temperature, we further lowered the temperature down to 300 mK. Figure 3(c) shows $R_{4pt}$ vs. T measurement of a Fe-doped $Bi_2Se_3$ nanoribbon device in the temperature regime down to 300 mK. Here, we find that the resistance increases logarithmically with temperature in the low temperature regime before saturating at 400 mK, characteristic features of the Kondo effect associated with magnetic impurities.[26] From the observed Kondo effect, we confirm that the VLS-grown $Bi_2Se_3$ nanoribbon using a Fe-Au thin film as a catalyst is magnetically doped with Fe. Identifying a precise Kondo temperature is complicated by lack of knowledge of the magnetic impurity concentration, by the fact that saturation in resistance is observed only slightly above our base temperature, and by the interesting structure of the low-temperature logarithmic dependence, which may even point to two different Kondo temperatures for Fe atoms on different sites.

We also measured magnetoresistances of the Fe-doped $Bi_2Se_3$ nanoribbons by sweeping magnetic fields perpendicular and parallel to the ribbons. In contrast to a clear weak antilocalization (WAL) effect observed in the un-doped $Bi_2Se_3$ nanoribbons reported recently,[16] we do not observe a clear WAL with Fe-doped $Bi_2Se_3$ nanoribbons. With un-doped $Bi_2Se_3$ nanoribbons, the WAL effect is expected due to the strong spin-orbit coupling[27]. In the case of Fe-doped $Bi_2Se_3$ nanoribbons, we attribute the lack of the WAL effect to the presence of the magnetic impurities, which can suppress the WAL effect by introducing spin flip scattering.[28]

From the low-temperature transport measurements, we confirmed that the $Bi_2Se_3$ nanoribbons are doped with Fe magnetic impurities. What would happen to the surface conducting states in the presence of the magnetic impurities? A theoretical prediction is that a small gap will open in the surface states due to the breaking of time-reversal symmetry.[17] To attempt to test this prediction, we can monitor the four-point resistance of the nanoribbon while applying a gate voltage which will sweep the position of the Fermi level through the band structure of the $Bi_2Se_3$ nanoribbon. In order to isolate the role of surface states, it would be necessary to minimize contribution from the residual bulk carriers to the overall transport signals.

Unfortunately, despite the enhanced surface-to-volume ratio of the $Bi_2Se_3$ nanoribbons, the surface effects are presently masked by the bulk carriers, which are the dominant transport channels in our ribbons. This is because the ribbons are a heavily doped n-type semiconductor, likely due to Se vacancies introduced during nanoribbon growth as well as exposure of the ribbon surface to air during processing and sample mounting.[29, 30] Figure 4 shows the carrier concentration of one of the Fe-doped $Bi_2Se_3$ nanoribbons by measuring the Hall resistance of the ribbon in the presence of a perpendicular magnetic field. First, we confirm the presence of Fe dopants in the ribbon by noting the increase in $R_{4pt}$ in the low temperature regime, as shown in Fig. 4(b). From Hall measurements, where a voltage drop perpendicular to the direction of the current flow is measured while a magnetic field is applied perpendicular to the plane of the ribbon surface (Fig. 4(a) schematic), we can estimate the carrier concentration. Figure 4(c) shows the Hall resistance, $R_H$, as a function of the applied magnetic field. The Hall slope is different between low and high magnetic fields as indicated by two dotted linear fits (red and black). The change in the Hall slope can arise from two factors: a different number of transport channels being introduced as a function of the magnetic field, and anomalous Hall effect introduced by Fe dopants and/or strong spin-orbit coupling in the nanoribbons, as in a recent experiment on a magnetically-doped 2D electron gas.[31] To separate these two contributing factors to the change in the Hall slope, temperature-dependent Hall measurements as well as Shubnikov-de Haas (SdH) oscillation measurements for independent carrier concentration calculations are needed. For the scope of the present work, we use the Hall slope for a rough estimate of carrier concentration. Using the Hall slope at low magnetic field and ignoring any potential contributions from the anomalous Hall effect, we estimate the carrier concentration per unit area to be ~ $1.2 \times 10^{14}$ cm$^{-2}$ (Assuming the ribbon thickness of 50 nm, which is the typical thickness of our nanoribbons, the bulk carrier volume density is ~ $2.3 \times 10^{19}$ cm$^{-3}$.) The areal carrier density we obtain here is at least one order of magnitude higher than the carrier concentration expected for the surface states.[16] Therefore, the transport measurements we obtain from the Fe-doped Bi$_2$Se$_3$ nanoribbons are dominated by bulk carriers. Figure 4(d) shows the oscillatory behavior of the Hall resistance as a function of the inverse magnetic field, after removing the linear dependence of the magnetic field in the Hall resistance. The Fourier-transform power spectrum of the Hall resistance is shown in the inset in Fig. 4(d). Only one dominant peak with the FFT frequency of ~ 67 T is observed, which we attribute to the bulk carrier channel. In the case of Bi$_2$Se$_3$ nanoribbons with no magnetic doping, some of the present authors reported a FFT frequency of ~ 71 T, assigning this to bulk carriers with concentration of ~ $1 \times 10^{19}$ cm$^{-3}$.[16] The difference in the bulk carrier concentrations between the Fe-doped and un-doped ribbons is not significant as it falls within the ribbon-to-ribbon variation.

We demonstrated how to dope magnetic species in Bi$_2$Se$_3$ nanoribbons *in-situ* and investigated the effect of magnetic doping on transport properties of Bi$_2$Se$_3$ nanoribbons. By using Fe-Au and Ni-Au thin films as catalysts for the VLS growth, we can magnetically dope the Bi$_2$Se$_3$ nanoribbons. The presence of the magnetic dopants in the nanoribbons is confirmed by the observation of the Kondo effect. Bulk carrier concentration is still too high in these magnetically doped Bi$_2$Se$_3$ nanoribbons to test whether a small gap opens in the surface states. Future work includes working on reducing the bulk carrier concentration in these $Bi_2Se_3$ nanoribbons and correlating the amount of Fe atoms in the ribbons with the carrier density.

**Acknowledgement.** . This work was made possible by support from the King Abdullah University of Science and Technology (KAUST) Investigator Award (No. KUS-l1-001-12) and by an NSF-NRI Supplement to the Center for Probing the Nanoscale, an NSF Nanoscale Science and Engineering Center (PHY-0425897). D. G.-G. acknowledges support from the D. and L. Packard Foundation and the Hellman Faculty Scholar program. J. R. W. acknowledges support from the K. van Bibber Postdoctoral Fellowship of the Stanford Physics Department. A. J. B. acknowledges support from an NDSEG Graduate Fellowship. P. G. acknowledges support from the Stanford Vice Provost for Undergraduate Education.

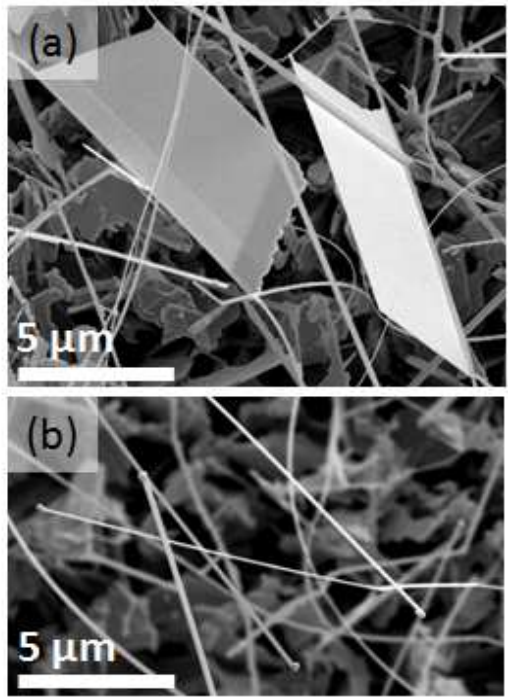

**Figure 1.** Bi$_2$Se$_3$ nanoribbons grown by the VLS mechanism using a 2 nm Fe/ 5 nm Au thin film as a catalyst. Two distinct morphologies are observed: (a) wide nanoribbons and (b) narrow, long nanoribbons. Catalyst particles are observed at the end of the long nanoribbons as shown in (b), suggesting that the nanoribbons are grown via the VLS mechanism.

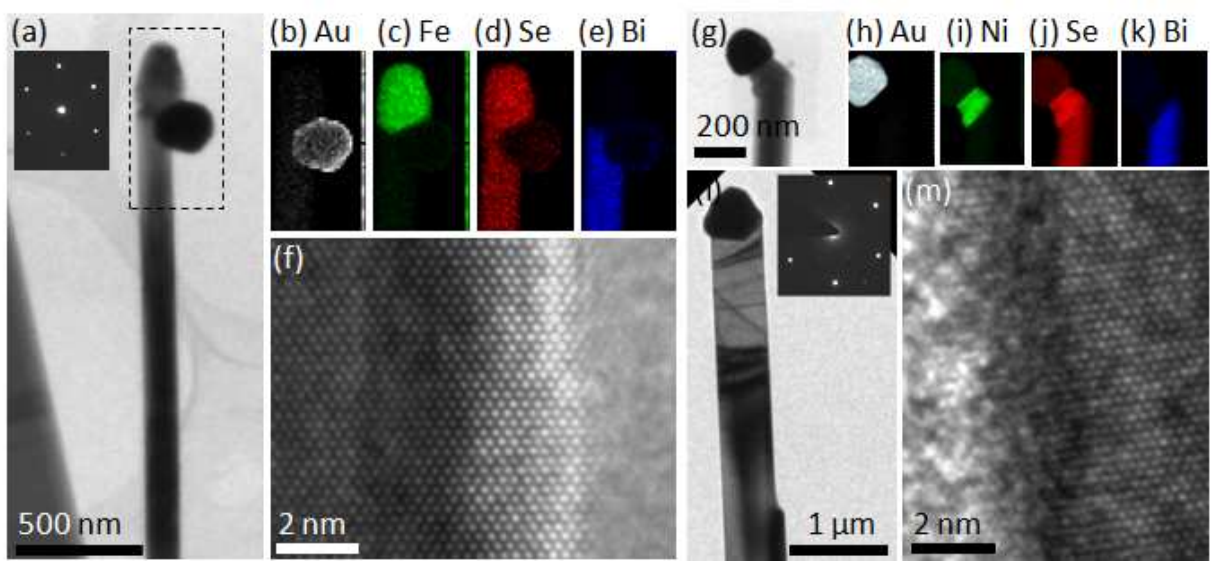

**Figure 2.** High-resolution TEM characterizations and EDX chemical maps of Bi$_2$Se$_3$ nanoribbons with Au-Fe-Se and Au-Ni-Se catalyst particles. (a) shows a Bi$_2$Se$_3$ nanoribbon grown using a Fe-Au thin film as a catalyst. The selected area diffraction pattern (inset in (a)) acquired from the ribbon confirms the high-quality, single-crystal nature of the Bi$_2$Se$_3$ nanoribbon. A 32x60 pixel EDX scan with an acquisition time of 500 ms per spectra was taken from the boxed region in (a) for chemical identification of the metal particle. (b-e) show Au, Fe, Se, and Bi maps obtained from the EDX scan respectively. (f) is a high-resolution TEM image of the ribbon in (a), showing the correct lattice spacing of ~ 0.21 nm. (g) shows a Bi$_2$Se$_3$ nanoribbon grown using a Ni-Au thin film. (h-k) show Au, Ni, Se, and Bi chemical maps obtained from a 50x70 pixel EDX scan that was taken from the ribbon shown in (g). A clear separation of Au and Ni-Se in the metal particle is observed from the chemical maps. (l) shows another Bi$_2$Se$_3$ nanoribbon grown with the Ni-Au thin film. The selected area diffraction pattern of the ribbon is shown in the inset in (l) and the lattice fringes with the correct lattice spacing are shown in (m).

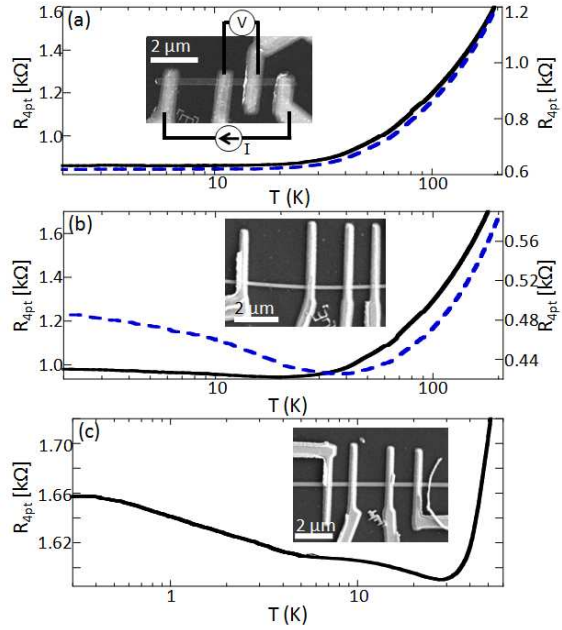

**Figure 3.** Temperature-dependent resistance measurements of un-doped and Fe-doped $Bi_2Se_3$ nanoribbons. (a) Four-point resistances ($R_{4pt}$) of two un-doped $Bi_2Se_3$ nanoribbons down to 2 K. The vertical axis on the left (right) corresponds to the curve in solid black (dotted blue). The inset shows a schematic of the four-point resistance set-up and the curve in solid black was acquired from the device shown in the inset. (b) Temperature-dependent resistances of two Fe-doped $Bi_2Se_3$ nanoribbons down to 2K, where the resistances increase with decreasing temperature below ~ 30 K. The device shown in the inset corresponds to curve in solid black. The vertical axis on the left (right) corresponds to the curve in solid black (dotted blue). (c) shows $R_{4pt}$ of another Fe-doped $Bi_2Se_3$ nanoribbon down to 300 mK. The logarithmic increase of the resistance on temperature below ~ 30 K indicates Kondo effect, confirming the presence of Fe dopants in the nanoribbons. The inset shows the corresponding nanoribbon device.

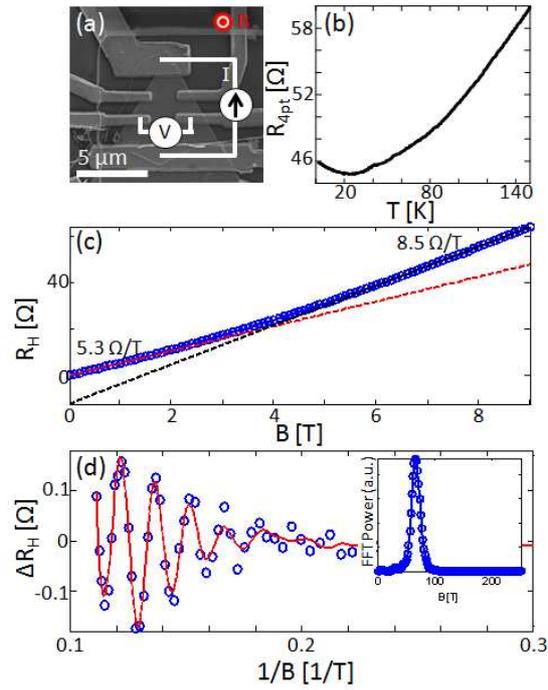

**Figure 4.** Hall resistance ($R_H$) of a Fe-doped $Bi_2Se_3$ nanoribbon for measuring a carrier concentration. The Fe-doped $Bi_2Se_3$ nanoribbon with Hall configuration electrodes is shown in (a). The schematic in (a) illustrates the Hall measurement set-up. The four-point resistance of the nanoribbon as a function of temperature down to 2K is shown in (b), showing the Kondo effect. (c) Hall resistance as a function of a magnetic field, applied perpendicular to the ribbon. Two slopes are extracted from the Hall resistance. The red dotted curve is a linear fit to $R_H$ in the low magnetic field while the black dotted curve is a linear fit to $R_H$ in the high magnetic field. (d) shows the oscillations in the Hall resistance, after removing the linear dependence in the resistance on the perpendicular magnetic field, as a function of inverse of the magnetic field. The blue dots represent data points, while the red curve is a fit to the data. The inset shows the power spectrum of the oscillations in the Hall resistance.